# Sterile Neutrinos and Future Solar Neutrino Experiments

S.M. Bilenky[a,b,c]* and C. Giunti[b,c]⋆

(a) Joint Institute of Nuclear Research, Dubna, Russia
(b) INFN Torino, Via P. Giuria 1, I–10125 Torino, Italy
(c) Dipartimento di Fisica Teorica, Università di Torino

## Abstract

It is shown that future solar neutrino experiments (SNO, Super-Kamiokande and others), in which high energy neutrinos will be detected (mostly from $^8$B decay), may allow to answer in a model independent way the question whether there are transitions of solar $\nu_e$'s into sterile states. No assumptions about the initial flux of $^8$B neutrinos are done. Lower bounds for the probability of transition of solar $\nu_e$'s into all possible sterile states are derived and expressed through measurable quantities.

\* BILENKY@TO.INFN.IT
⋆ GIUNTI@TO.INFN.IT

Solar neutrino experiments are very important sources of informations about neutrino properties (masses, mixing, magnetic moment, etc.) and about the central invisible part of the sun, where energy is produced. As is well known [1–4], the existing solar neutrino data can be described by the resonant MSW enhancement in matter of the mixing between two neutrino flavours [1]. For the parameters $\Delta m^2 = m_2^2 - m_1^2$ and $\sin^2 2\vartheta$ ($m_1$, $m_2$ are neutrino masses, and $\vartheta$ is the vacuum mixing angle) a small mixing angle solution ($\Delta m^2 \simeq 5 \times 10^{-6}$ eV$^2$ and $\sin^2 2\vartheta \simeq 8 \times 10^{-3}$) and a large mixing angle solution ($\Delta m^2 \simeq 10^{-5}$ eV$^2$ and $\sin^2 2\vartheta \simeq 0.8$) were found. Let us stress, however, that these solutions were obtained under the assumption that the Standard Solar Model (SSM) [7, 8] correctly predicts the neutrino fluxes from different reactions of the $pp$ and CNO thermonuclear cycles.

A new generation of real-time high-statistics solar neutrino experiments (SNO [9] and Super-Kamiokande [10]) is currently in preparation. The feasibility of other proposed experiments (BOREXINO [11], ICARUS [12], HELLAZ [13] and others) is under investigation. In these experiments solar neutrinos will be detected by the observation of charged-current processes (CC), as well as neutral current processes (NC) and elastic neutrino-electron scattering (ES). The possibility of a model independent treatment of future solar neutrino data were discussed in ref.[14, 15]. In ref.[15] we have shown that the data that will be obtained by the SNO and Super-Kamiokande experiments (scheduled to start in 1995-96), in which mostly high energy $^8$B neutrinos will be detected, will allow to determine the survival probability $P_{\nu_e \to \nu_e}(E)$ as a function of neutrino energy $E$ independently from the total initial neutrino flux and the total initial neutrino flux $\Phi_{\nu_e}^{^8\text{B}}$ independently from what is going on with neutrinos on their way from the sun to the detector. This result was based on the assumption that solar $\nu_e$'s can transform only into other active neutrinos, i.e. $\nu_\mu$'s and/or $\nu_\tau$'s. However, in the general case of neutrino mixing active neutrinos can transform not only into other active neutrinos but also into sterile left-handed (anti)neutrinos that do not interact with matter via standard weak interactions (see ref.[16]). The problem of the existence of sterile neutrinos is of fundamental importance for the theory of neutrino mixing (the notion of sterile neutrinos was introduced by Pontecorvo [17]). The proof that there are transitions of active neutrinos into sterile states would suggest that 1) the neutrino mass term is of Dirac-Majorana type, 2) neutrinos with definite mass are Majorana particles and 3) the number of massive neutrinos is larger than the number of lepton flavours. Let us notice that a Dirac-Majorana mass term naturally appears in GUT models (see, for example, ref.[18]).

In the present paper we will show that future solar neutrino experiments may allow to answer in a model independent way the question whether there are transitions of solar $\nu_e$'s into sterile states. We will also obtain some model independent lower bounds for the transition probability of $\nu_e$'s into sterile states.

In the SNO experiment solar neutrinos will be detected through observation of the

---

[1] The existing solar neutrino data can also be described with other mechanisms: vacuum oscillations [4], neutrino decay [5], resonant spin-flavour precession [6] and others.

following three reactions:

$$\nu_e + d \to e^- + p + p \quad \text{(CC)} \tag{1}$$

$$\nu + d \to \nu + p + n \quad \text{(NC)} \tag{2}$$

$$\nu + e^- \to \nu + e^- \quad \text{(ES)} \tag{3}$$

Since the energy threshold in the SNO experiment will be rather high ($\gtrsim 5\,\text{MeV}$ for CC and ES and 2.2 MeV for NC), mostly $^8$B neutrinos will contribute to the event rates. Reaction (3) will be investigated in detail also by the Super-Kamiokande experiment with a threshold $\gtrsim 5\,\text{MeV}$. The energy spectrum of $^8$B neutrinos is given by

$$\phi_{\nu_e}^{^8\text{B}}(E) = \Phi_{\nu_e}^{^8\text{B}} X(E) \tag{4}$$

Here the function $X(E)$ is the normalized neutrino spectrum from the decay $^8\text{B} \to {}^8\text{Be} + e^+ + \nu_e$, which is determined by the phase space factor (corrections due to forbidden transitions where calculated in ref.[19]). The distortions of the neutrino spectra are negligibly small under solar conditions [20].

Since sterile neutrinos are not detected, information about their existence can be obtained only from the conservation of probability

$$\sum_{\ell=e,\mu,\tau} P_{\nu_e \to \nu_\ell}(E) + P_{\nu_e \to \nu_S}(E) = 1 \,, \tag{5}$$

where the two terms in the left hand side are the probability of transition of solar $\nu_e$'s into all possible active states and all possible left-handed sterile states (quanta of right-handed fields), respectively. At first sight it could seem that it is not possible to derive relations which may allow us to reveal the existence of sterile neutrinos without any assumption about the value of the initial neutrino flux. However, this is possible if solar neutrinos are detected by the observation of **different reactions**. In ref.[15], using the fact that the same initial neutrino flux $\Phi_{\nu_e}^{^8\text{B}}$ can be connected with the NC event rate as well as the ES and CC event rates, we obtained the following relation:

$$N^{\text{NC}} - r\,\Sigma^{\text{ES}} = r \int_{E_{\text{th}}} \sigma_{\nu_\mu e}(E)\,\phi_{\nu_S}(E)\,\text{d}E - \int_{E_{\text{th}}} \sigma_{\nu d}^{\text{NC}}(E)\,\phi_{\nu_S}(E)\,\text{d}E \,, \tag{6}$$

where

$$\Sigma^{\text{ES}} \equiv N^{\text{ES}} - \int_{E_{\text{th}}} \left( \sigma_{\nu_e e}(E) - \sigma_{\nu_\mu e}(E) \right) \phi_{\nu_e}(E)\,\text{d}E \,, \tag{7}$$

$N^{\text{NC}}$ and $N^{\text{ES}}$ are the total NC and ES event rates,

$$\phi_{\nu_e}(E) = \frac{n^{\text{CC}}(E)}{\sigma_{\nu_e d}^{\text{CC}}(E)} \,, \tag{8}$$

is the $\nu_e$ flux on the earth which will be measured by detecting the CC process (1) ($n^{\text{CC}}(E)$ is the differential CC event rate) and $\phi_{\nu_S}(E)$ is the flux of sterile neutrinos on the earth. The quantity $r$ in Eq.(6) is defined as

$$r \equiv \frac{X^{\text{NC}}_{\nu d}}{X^{\text{ES}}_{\nu_\mu e}} \tag{9}$$

where

$$X^{\text{NC}}_{\nu d} \equiv \int_{E_{\text{th}}} \sigma^{\text{NC}}_{\nu d}(E) \, X(E) \, \mathrm{d}E \tag{10}$$

$$X^{\text{ES}}_{\nu_\mu e} \equiv \int_{E_{\text{th}}} \sigma_{\nu_\mu e}(E) \, X(E) \, \mathrm{d}E \tag{11}$$

are known quantities: $X^{\text{NC}}_{\nu d} = 4.1 \times 10^{-43} \, \text{cm}^2$ [21] and $X^{\text{ES}}_{\nu_\mu e} = 2.7 \times 10^{-45} \, \text{cm}^2$ [22], which gives $r = 1.5 \times 10^2$; $\sigma^{\text{CC}}_{\nu_e d}(E)$, $\sigma^{\text{NC}}_{\nu d}(E)$ and $\sigma_{\nu_\mu e}(E)$ are the cross sections for the CC, NC and $\nu_\mu \, e^- \to \nu_\mu \, e^-$ reactions, respectively.

Eq.(6) could allow to check in a model independent way whether there are transitions of solar $\nu_e$ into sterile neutrinos. In fact, if the left hand part of Eq.(6), in which only measurable quantities enter, is not equal to zero, it would mean that there are sterile neutrinos in the solar neutrino flux on the earth. Two comments are in order: 1) According to our model calculation the two terms in the right-hand side of Eq.(6) could cancel each other. 2) Eq.(6) cannot give any information about the value of the transition probability of $\nu_e$'s into sterile neutrinos, even if the presence of sterile neutrinos will be revealed.

In the present paper we will derive different relations which will allow to check in a model independent way whether there are transitions of solar $\nu_e$'s into sterile states. For the probability $P_{\nu_e \to \nu_S}$ we will derive lower bounds and express them through measurable quantities.

Let us consider the NC process (2). Using $\nu_e$-$\nu_\mu$-$\nu_\tau$ universality of neutral current, for the average value of the transition probability of $\nu_e$'s into all types of active neutrinos we have

$$\left\langle \sum_{\ell=e,\mu,\tau} P_{\nu_e \to \nu_\ell} \right\rangle_{\text{NC}} = \frac{N^{\text{NC}}}{X^{\text{NC}}_{\nu d} \, \Phi^{8\text{B}}_{\nu_e}} \tag{12}$$

where

$$\left\langle \sum_{\ell=e,\mu,\tau} P_{\nu_e \to \nu_\ell} \right\rangle_{\text{NC}} \equiv \frac{\int_{E_{\text{th}}} \sigma^{\text{NC}}_{\nu d}(E) \, X(E) \sum_{\ell=e,\mu,\tau} P_{\nu_e \to \nu_\ell}(E) \, \mathrm{d}E}{X^{\text{NC}}_{\nu d}} \tag{13}$$

If it occurs that $\left\langle \sum_{\ell=e,\mu,\tau} P_{\nu_e \to \nu_\ell} \right\rangle_{\text{NC}}$ is less than one, it would mean that sterile neutrinos exist. However, as it is clear from Eq.(12), without the knowledge of the total neutrino flux it is impossible to reach any conclusion about the average probability $\left\langle \sum_{\ell=e,\mu,\tau} P_{\nu_e \to \nu_\ell} \right\rangle_{\text{NC}}$.
Let us turn to the measurement of the CC event rate. From this measurement it is possible to obtain a model independent **lower bound** for the total neutrino flux. In fact we have

$$P_{\nu_e \to \nu_e}(E) = \frac{\phi_{\nu_e}(E)}{\Phi_{\nu_e}^{^8\text{B}} X(E)} \quad (14)$$

where the flux $\phi_{\nu_e}(E)$ is given by Eq.(8). Since $P_{\nu_e \to \nu_e}(E) \leq 1$, from Eq.(14) we obtain

$$\Phi_{\nu_e}^{^8\text{B}} \geq (\phi_{\nu_e}/X)_{\max} \quad (15)$$

where the subscript max indicates the maximum value of $\phi_{\nu_e}(E)/X(E)$ in the explored energy range. From Eq.(12) and Eq.(15) we have

$$\left\langle \sum_{\ell=e,\mu,\tau} P_{\nu_e \to \nu_\ell} \right\rangle_{\text{NC}} \leq \mathcal{R}^{\text{NC}} \quad (16)$$

where

$$\mathcal{R}^{\text{NC}} \equiv \frac{N^{\text{NC}}}{X_{\nu d}^{\text{NC}} (\phi_{\nu_e}/X)_{\max}} \quad (17)$$

is a measurable quantity (through the observation of NC and CC event rates). If it occurs that $\mathcal{R}^{\text{NC}} < 1$ it would mean that sterile neutrinos exist [2]. In this case, for the average transition probability of solar $\nu_e$'s into all possible sterile states we have

$$\langle P_{\nu_e \to \nu_S} \rangle_{\text{NC}} \geq 1 - \mathcal{R}^{\text{NC}} \quad (18)$$

Let us discuss now the possibility to reveal the presence of sterile neutrinos in the solar neutrino flux on the earth from the measurement of the total ES event rate and the CC event rate. In this case we have

$$\left\langle \sum_{\ell=e,\mu,\tau} P_{\nu_e \to \nu_\ell} \right\rangle_{\text{ES}} = \frac{\Sigma^{\text{ES}}}{X_{\nu_\mu d}^{\text{ES}} \Phi_{\nu_e}^{^8\text{B}}} \quad (19)$$

where

$$\left\langle \sum_{\ell=e,\mu,\tau} P_{\nu_e \to \nu_\ell} \right\rangle_{\text{ES}} \equiv \frac{\int_{E_{\text{th}}} \sigma_{\nu_\mu e}(E) X(E) \sum_{\ell=e,\mu,\tau} P_{\nu_e \to \nu_\ell}(E) \, \mathrm{d}E}{X_{\nu_\mu e}^{\text{ES}}} \quad (20)$$

---
[2] It is clear that in the case $\mathcal{R}^{\text{NC}} \geq 1$ no conclusion on the existence of sterile neutrinos can be reached.

and $\Sigma^{\text{ES}}$ is given by Eq.(7). Using Eq.(15) and Eq.(19) we obtain

$$\left\langle \sum_{\ell=e,\mu,\tau} P_{\nu_e \to \nu_\ell} \right\rangle_{\text{ES}} \leq \mathcal{R}^{\text{ES}} \tag{21}$$

where

$$\mathcal{R}^{\text{ES}} \equiv \frac{\Sigma^{\text{ES}}}{X^{\text{ES}}_{\nu_\mu e} \, (\phi_{\nu_e}/X)_{\max}} \tag{22}$$

is a measurable quantity (through the observation of ES and CC event rates). If it occurs that $\mathcal{R}^{\text{ES}} < 1$ it would mean that solar $\nu_e$'s transform into sterile states and the average transition probability of solar $\nu_e$'s into all possible sterile states have the following lower bound:

$$\langle P_{\nu_e \to \nu_S} \rangle_{\text{ES}} \geq 1 - \mathcal{R}^{\text{ES}} \tag{23}$$

Let us return now to Eq.(17) and Eq.(22). It is easy to show that

$$\mathcal{R}^{\text{NC;ES}} = \frac{\left\langle \sum_{\ell=e,\mu,\tau} P_{\nu_e \to \nu_\ell} \right\rangle_{\text{NC;ES}}}{P^{\max}_{\nu_e \to \nu_e}} \tag{24}$$

where $P^{\max}_{\nu_e \to \nu_e} = \frac{1}{\Phi^{^8\text{B}}_{\nu_e}} (\phi_{\nu_e}/X)_{\max}$. Three comments are in order: 1) The closer $P^{\max}_{\nu_e \to \nu_e}$ is to one, the closer $\left\langle \sum_{\ell=e,\mu,\tau} P_{\nu_e \to \nu_\ell} \right\rangle_{\text{NC;ES}}$ is to $\mathcal{R}^{\text{NC;ES}}$. 2) Each probability in the right hand side of Eq.(24) depends on $\Phi^{^8\text{B}}_{\nu_e}$. However, the ratios $\mathcal{R}^{\text{NC}}$ and $\mathcal{R}^{\text{ES}}$ do not depend on $\Phi^{^8\text{B}}_{\nu_e}$. This is the basis of the model independent tests discussed here. 3) From Eq.(24) it is obvious that if the sum $\sum_{\ell=e,\mu,\tau} P_{\nu_e \to \nu_\ell}$ does not depend on energy then $\mathcal{R}^{\text{NC;ES}} \geq 1$. No conclusion about the existence of sterile neutrinos can be reached in this case. Thus, if it occurs that $\mathcal{R}^{\text{NC}} < 1$ or $\mathcal{R}^{\text{ES}} < 1$ it would mean not only that sterile neutrinos exist but also that $P_{\nu_e \to \nu_S}$ depends on neutrino energy $E$.

In Table 1 the result of calculations of $P^{\max}_{\nu_e \to \nu_e}$ and lower bounds for $\langle P_{\nu_e \to \nu_S} \rangle_{\text{NC;ES}}$ in the simplest model with transitions of solar $\nu_e$'s into a sterile state $\nu_S$ are presented. In the calculation we used values of the parameters $\Delta m^2$ and $\sin^2 2\vartheta$ that where found in ref.[2, 4] from the fit of the existing data under the assumption that the initial neutrino flux is given by the SSM [3]. From Table 1 it can be seen that the values of the ratios $\mathcal{R}^{\text{NC;ES}}$ that characterize the presence of sterile neutrinos are less than one for all the values of $\Delta m^2$ and $\sin^2 2\vartheta$ considered. This means that the tests of sterility discussed here may yield significant results if sterile neutrinos exist.

---

[3] Let us remark that in this model there is no large mixing angle solution.

In Super-Kamiokande and other future solar neutrino experiments (ICARUS, etc.) a large number of solar neutrino induced ES events will be observed. From these data the differential ES event rate $n^{\mathrm{ES}}(E)$ will be determined and new possibilities for testing the existence of sterile neutrinos will emerge. In fact, a measurement of $n^{\mathrm{ES}}(E)$ and $n^{\mathrm{CC}}(E)$ will allow to determine the differential flux of all types of active neutrinos on the earth:

$$\sum_{\ell=e,\mu,\tau} \phi_{\nu_\ell}(E) = \frac{n^{\mathrm{ES}}(E)}{\sigma_{\nu_\mu e}(E)} + \left(1 - \frac{\sigma_{\nu_e e}(E)}{\sigma_{\nu_\mu e}(E)}\right) \frac{n^{\mathrm{CC}}(E)}{\sigma_{\nu_e d}^{\mathrm{CC}}(E)} . \tag{25}$$

From Eq.(5) we obtain

$$\frac{1}{X(E)} \sum_{\ell=e,\mu,\tau} \phi_{\nu_\ell}(E) = \Phi_{\nu_e}^{^8\mathrm{B}} \left[1 - P_{\nu_e \to \nu_\mathrm{S}}(E)\right] . \tag{26}$$

If it will occur that the left-hand side of this equation, which contains only measurable quantities, depends on energy, then it would mean that there are sterile neutrinos in the flux of solar neutrinos on the earth. Furthermore, in the case under consideration we can obtain an upper bound for the transition probability of solar $\nu_e$'s into all types of active neutrinos as a function of energy:

$$\sum_{\ell=e,\mu,\tau} P_{\nu_e \to \nu_\ell}(E) \leq \mathrm{r}^{\mathrm{ES}}(E) \tag{27}$$

where

$$\mathrm{r}^{\mathrm{ES}}(E) \equiv \frac{\displaystyle\sum_{\ell=e,\mu,\tau} \phi_{\nu_\ell}(E)}{X(E) \left(\displaystyle\sum_{\ell=e,\mu,\tau} \phi_{\nu_\ell}/X\right)_{\mathrm{max}}} \tag{28}$$

From Eq.(27), we obtain the following lower bound for the transition probability of solar $\nu_e$'s into sterile states as a function of energy [4]:

$$P_{\nu_e \to \nu_\mathrm{S}}(E) \geq 1 - \mathrm{r}^{\mathrm{ES}}(E) \tag{29}$$

In Figure 1 the survival probability $P_{\nu_e \to \nu_e}(E)$ and the lower bound for the transition probability $P_{\nu_e \to \nu_\mathrm{S}}(E)$ in the simplest model with $\nu_e \to \nu_\mathrm{S}$ transitions are depicted. The used values of the parameters $\Delta m^2$ and $\sin^2 2\vartheta$ are given in Table 1. It can be seen from Figure 1 that the lower bounds for $P_{\nu_e \to \nu_\mathrm{S}}(E)$ have a strong energy dependence in the model considered. Different values of the parameters $\Delta m^2$ and $\sin^2 2\vartheta$ give completely different behaviours of the lower bound.

In conclusion let us make the following remarks:

---

[4] If $\sum_{\ell=e,\mu,\tau} \phi_{\nu_\ell}(E)/X(E)$ is energy independent, then $\mathrm{r}^{\mathrm{ES}}(E) = 1$ and no conclusion on the existence of sterile neutrinos can be reached.

1) The inequality (15) could allow us to test the SSM in a model independent way. In fact, if it occurs that $\Phi_{\nu_e}^{^8B}(\text{SSM}) < (\phi_{\nu_e}/X)_{\text{max}}$, it would mean that the solar neutrino flux predicted by the SSM is lower than the real flux [5].

2) If the transition probability of solar $\nu_e$'s into sterile states is energy independent, no model independent information on sterile neutrinos can be obtained, as can be seen from Eq.(26).

3) We assumed here that there are no active antineutrinos in the solar neutrino flux on the earth (due to possible neutrino magnetic moment effects, neutrino decay, etc.). This case will be discussed elsewhere.

---

[5] This possibility seems rather unusual. Notice, however, that it could be realized in the case of a strong energy dependence of the survival probability.

| $\nu_e \to \nu_S$ | $\Delta m^2$ (eV$^2$) | $\sin^2 2\vartheta$ | $P^{\max}_{\nu_e \to \nu_e}$ | Lower Bound for | |
|---|---|---|---|---|---|
| | | | | $\langle P_{\nu_e \to \nu_S} \rangle_{\text{NC}}$ | $\langle P_{\nu_e \to \nu_S} \rangle_{\text{ES}}$ |
| VACUUM OSC. (a) | $9.2 \times 10^{-11}$ | 0.85 | 0.85 | 0.61 | 0.70 |
| VACUUM OSC. (b) | $6.4 \times 10^{-11}$ | 0.8 | 0.59 | 0.46 | 0.47 |
| SMALL MIX MSW | $4.1 \times 10^{-6}$ | $9.1 \times 10^{-3}$ | 0.52 | 0.34 | 0.29 |

Table 1: Maximum values of the survival probability in the energy range that will be explored by SNO with $T_e^{\text{th}} = 5.0$ MeV and corresponding values of the lower bounds for the average probabilities of transition of solar $\nu_e$'s into sterile neutrinos $\langle P_{\nu_e \to \nu_S} \rangle_{\text{NC;ES}}$ for three possible solutions of the solar neutrino problem with transitions of $\nu_e$'s into $\nu_S$ [2, 4].

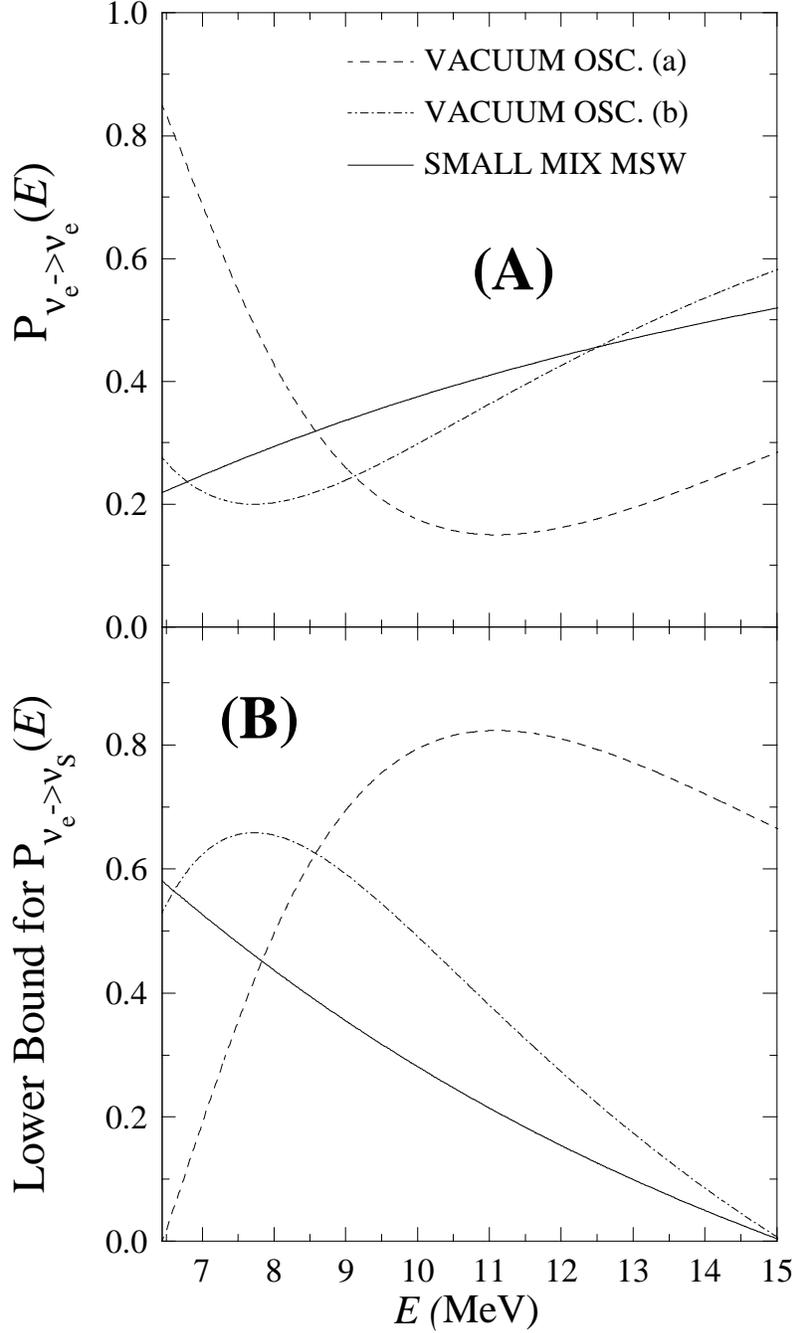

Figure 1A: Survival probability for three possible solutions of the solar neutrino problem with transitions of solar $\nu_e$'s into sterile neutrinos. Figure 1B: Corresponding lower bound for the transition probability of solar $\nu_e$'s into sterile neutrinos. The depicted energy range will be explored by SNO with $T_e^{\text{th}} = 5.0\,\text{MeV}$. The parameters $\Delta m^2$ and $\sin^2 2\vartheta$ are given in Table 1.